\documentclass[12pt]{article} 
\usepackage[sectionbib]{natbib}
\usepackage{array,epsfig,fancyheadings,rotating}
\usepackage[]{hyperref}  
\usepackage{amsmath}
\usepackage{cleveref}
\usepackage{sectsty, secdot}
\sectionfont{\fontsize{12}{14pt plus.8pt minus .6pt}\selectfont}
\renewcommand{\theequation}{\thesection\arabic{equation}}
\subsectionfont{\fontsize{12}{14pt plus.8pt minus .6pt}\selectfont}

\usepackage{amsmath}
\usepackage{amssymb}
\usepackage{amsfonts}
\usepackage{multirow}
\usepackage{amsthm}
\usepackage{enumerate}
\usepackage{threeparttable}
\usepackage{multirow}
\usepackage[small]{caption}
\usepackage{booktabs}

\newtheorem{theorem}{Theorem}

\theoremstyle{definition}

\newtheorem{remark}{Remark}
\def\argmax{\mathop{\rm argmax}}
\def\argmin{\mathop{\rm argmin}}
\def\Var{\mathop{\rm Var}}

\begin{document}


\renewcommand{\baselinestretch}{2}

\markboth{\hfill{\footnotesize\rm Miaomiao Su AND Qihua Wang} \hfill}
{\hfill {\footnotesize\rm Adjusted Estimating Equation Estimation for Quantile} \hfill}

$\ $\par


\fontsize{12}{10pt plus.8pt minus .6pt}\selectfont \vspace{0.8pc}
\centerline{\large\bf A Convex Programming Solution Based 
Debiased Estimator }
\vspace{2pt} 
\centerline{\large\bf  for Quantile with Missing Response and}
\centerline{\large\bf High-dimensional Covariables} 
\centerline{Miaomiao Su and Qihua Wang} 
\centerline{\it
Academy of Mathematics and Systems Science, 
Chinese Academy of Sciences
}
\centerline{\it
University of Chinese Academy of Sciences
}
\centerline{E-mail: smm@amss.ac.cn;\quad  qhwang@amss.ac.ac.cn}
 \vspace{.55cm} \fontsize{9}{11.5pt plus.8pt minus.6pt}\selectfont


\begin{quotation}
\noindent {\it Abstract:}
{\bf }\\
This paper is concerned with the estimating problem of response quantile with high dimensional covariates when response is missing at random.
Some existing methods define root-$n$ consistent estimators for the response quantile. But these methods require correct specifications of both the conditional distribution of response given covariates and the selection probability function. 
 In this paper, a debiased method is proposed  by solving a convex programming. The estimator obtained by the proposed method is asymptotically normal given a correctly specified parametric model for the condition distribution function, without the requirement to specify the selection probability function. Moreover, the proposed estimator can be asymptotically more efficient than the existing estimators. The proposed method is evaluated by a simulation study and is illustrated by a real data example.

\vspace{9pt}
\par
\end{quotation}\par

\def\thefigure{\arabic{figure}}
\def\thetable{\arabic{table}}

\renewcommand{\theequation}{\thesection.\arabic{equation}}

\fontsize{12}{14pt plus.8pt minus .6pt}\selectfont

\lhead[\footnotesize\thepage\fancyplain{}\leftmark]{}\rhead[]{\fancyplain{}\rightmark\footnotesize\thepage}

\section{Introduction}

The estimation and inference problem with missing responses is an important topic in statistics and has been studied extensively. 
It may define a biased estimator and lead to a loss of efficiency by simply ignoring the subjects with missing responses. 
 This inspires the development of some approaches, including the imputation, inverse probability weighting and doubly robust methods. 
 See, for example, \cite{rosenbaum1983central}, \cite{hahn1998role}, \cite{Hirano2003efficient}, \cite{cao2009improving}, \cite{Rotnitzky2012improved}, \cite{Firpo2007Quantile}, \cite{Wang2010Quantile}, \cite{Hu2011Quantile} , \cite{Zhang2011Quantile}, and \cite{Melly2013Quantile}. 
Many early literature established  asymptotic theory on estimation and inference problem with missing responses in the classical setting where the dimension of covariable vector is a constant. 
However, the asymptotic results established in the classical setting may not hold in the high dimensional setting when dimension $p$ of the covariable vector diverges with 
the sample size $n$ and even possibly is larger than $n$. 
  With the help of some important techniques in high dimension, such as the Lasso (\cite{Tibshirani1996Lasso}), adaptive-lasso (\cite{Zou2006adaptive}), elastic-net lasso (\cite{Zou2005elastic}),  there has been considerable recent developments on  estimation and inference problems when the response is missing at random with high dimensional covariates.
  
 There are  some important advances  for statistical inference on the response mean. 
  \cite{Farrell2015high} extended the augmented inverse probability weighting approach in classical setting to high dimensions by incorporating the regularized penalization to both the outcome regression model and the selection probability function simultaneously, and proposed an asymptotically normal estimator for the response mean. 
  Although the estimator proposed by \cite{Farrell2015high} can be used to make inference on the response mean, it crucially relies on correct specifications of both the outcome regression model and the selection probability function. 
  To alleviate the conditions on model specification of unknown functions, \cite{Athey2018debias} proposed an approximate residual balancing debiasing method and obtained a $\sqrt{n}$-consistent estimator for the response mean with a correctly specified linear model on the outcome regression model without the requirement to specify the selection probability function.  
 However, it is somewhat unfortunate that the asymptotic normality has not been 
 proved for the estimator. It should be mentioned that the debiasing techniques  are  also used by \cite{javanmard2014high}, \cite{Buhlmann2014high} and\cite{zhang2014high}) for estimating regression coefficients.
   
Another important issue is the estimation of the marginal response quantile. In this paper, we consider the estimation of the marginal response quantile with response missing at random and high dimensional covariate vector.
There are many researches focusing on quantile regression \citep{He2016QR,  Belloni2019Quantile, Pietrosanu2021QR, han2019quantile} with low dimensional or high dimensional covariate vector.
However, one can not resort to the conditional quantile regression to obtain an estimator for the marginal response quantile directly. 
To our knowledge, the only estimator for the marginal response quantile that is shown to be $\sqrt{n}$ asymptotically normal is proposed in \cite{Belloni2017Quantile}. 
The $\sqrt{n}$-consistency of the estimator in \cite{Belloni2017Quantile} needs that both the conditional distribution of the response given covariates and the selection probability function are correctly specified.  
This motivates us to propose a new method for estimation of the marginal response quantile. 
This method defines an asymptotically normal estimator for the response quantile in the setting where the condition distribution function is assumed to be  a correctly specified parametric model, without the  requirement  to specify the selection probability function.  Moreover, the proposed estimator can be asymptotically more efficient since the asymptotic variance of the proposed estimator is less than or equal to that in \cite{Belloni2017Quantile}. This method consists of 
 the following three steps. 
First, we assume a single index model for the conditional distribution of the response given covariates and establish a conditional distribution based estimating equation. 
Second, we make an adjustment on the equation by adding the difference between the weighted empirical distribution and  the weighted conditional distribution  to the estimating  function. 
Third, we solve the adjusted estimation equation to obtain the proposed estimator. 
The weight in the second step is obtained by solving a convex programming which 
makes  the variance of the proposed estimator attain minimum and
constrains its bias such that it is $\sqrt{n}$-consistent.
 
 The rest of this paper is organized as follows.  
 In \Cref{sec: 2}, we develop a debiased estimating method by solving a convex 
 programming. 
 All assumptions and asymptotic properties are stated in \Cref{sec: 3}. 
\Cref{sec: 4} provides an equivalent easy-to-implement   method for calculating the weights.
 \Cref{sec: 5} presents some simulation studies to examine the finite sample performance of the proposed method. The real data application is reported in \Cref{sec: 6}. Outlines of the proofs of the main theorems are presented in the Appendix and the technical details are relegated to the supplementary material.

\section{Methodology}\label{sec: 2}

We first introduce some frequently used notations. 
 For positive sequences $a_n$ and $b_n$, let $b_n\asymp a_n$ denote $\lim_{n\rightarrow \infty}a^{-1}_nb_n=c$ for a positive constant $c$. Let $b_n\gtrsim a_n$ denote $\lim_{n\rightarrow \infty}a^{-1}_nb_n\geq c$ and $b_n\lesssim a_n$ denote $\lim_{n\rightarrow \infty}a^{-1}_nb_n\leq c$.
For a $p$-dim vector $v$, we let $\left\|v\right\|_{0}$ denote the number of non-zero elements in $v$, $\left\|v\right\|_{\infty} = \max\{|v_1|,\cdots,|v_p|\}$, $\left\|v\right\|_q=\left(\sum^{p}_{i=1}|v_i|^{q}\right)^{1/q}$ for $1\leq q<\infty$ and $|v| = (|v_1|,\cdots,|v_p|)^{\top}$. 
For two $p$-dim vectors $u$ and $v$, let $u^{\top}v = \sum^p_{j=1}u_jv_j$ and $u\odot v=(u_1v_1,\cdots,u_pv_p)^{\top}$.
For a matrix $A\in \mathbb{R}^{m\times n}$, $\left\|A\right\|_{\infty}$ denotes the largest absolute value of the elements in $A$.
For a $p$-dim vector $v$ and a set $S\subseteq \{1,\cdots,p\}$, let $|S|$ denote the number of elements in $S$, $v_{S}$ the vector in which $v_{Sj}=v_j$ if $j\in S$, $v_{Sj}=0$ if $j\notin S$, and $v_{-S}$ the vector removing the elements of $v$ corresponding to the index in $S$.

Let $Y$ be the response variable and $X$ the $p$ dimension covariable vector. 
The population $\tau-$quantile of $Y$ is defined as
\[q_{0}(\tau) \equiv  \inf \{y: F_{Y}(y)\geq \tau\},\]
where $F_{Y}(\cdot)$ is the distribution function of $Y$ and $0<\tau < 1$ is a constant. 
For ease of notation, we write $q_0(\tau)$ to be $q_0$ hereafter. We consider the case where $X$ is always observed and $Y$ is missing. Let $\delta$ denote the binary missing indicator for $Y$; that is, $\delta=1$ if $Y$ is observed; otherwise, $\delta=0$. Throughout this paper, we assume that the response is missing at random, that is, $P(\delta=1 \mid Y, X) = P(\delta=1\mid X)$, a commonly used missing mechanism in literature of statistical inference with missing data (\cite{rosenbaum1983central}). Statistical inference on $q_0$ cannot proceed without further restrictions when $p$ diverges with $n$. 
Hence, we impose the following structure on the conditional density of response given covariates $f_{Y\mid X}(y\mid X=x) = f(y, x^{\top}\beta)$ with $f$ being a known function and $\beta$ a $p$-dim model parameter vector. The  true unknown parameter vector is denoted by $\beta^0$.
Although $p$ may be larger than $n$, it is often true that only $\|\beta^0\|_0$ among $p$ covariates have nonzero coefficients and $\|\beta^0\|_0$ is fixed or diverges much slower than $n$.
 The corresponding conditional distribution function is denoted by $h(y,x^{\top}\beta) = \int_{-\infty}^{y}f(u,x^{\top}\beta)du$. 
 Without loss of generality, we assume that the intercept is zero and all covariates are centered (See, for example, Section 2.2 in \cite{Buhlmann2011uniform}, page 8). 

  Suppose we have $n$ independent and identically distributed observations 
\[\left({X}_{i},\delta_{i},Y_{i}\right), i= 1,\cdots,n,\]
 where some of $Y_i$'s are missing and $X_i$'s are completely observed. It is noted that $E[h(y,X^{\top}\beta^0)]=F_{Y}(y)$. A natural way to estimate $q_{0}$ is to solve 
\begin{equation}\label{eq: imputation}
\frac{1}{n}\sum_{i=1}^{n}h(q,X^{\top}_{i}\hat{\beta})= \tau,
\end{equation}
 where $\hat{\beta}$ is given by the following Lasso method with complete case (CC) analysis
\begin{equation}\label{lasso beta}
\hat{\beta} = \argmin_{\beta}-\sum^{n}_{i=1}\delta_{i}\log(f(Y_{i},X^{\top}_{i}\beta)) + \lambda\left\|\beta\right\|_{1}
\end{equation}
and $\lambda\asymp \sqrt{\log(p)/n}$. The solution of \eqref{eq: imputation} is denoted by $\tilde{q}$. 
 $\tilde{q}$ is typically not a $\sqrt{n}$-consistent estimator of $q_{0}$ since $\hat{\beta}$ is generally not $\sqrt{n}$-consistent due to its high dimensionality. 
Therefore, to define a $\sqrt{n}$-consistent estimator for $q_{0}$, we consider a modified estimation equation of \eqref{eq: imputation}, which is given as follows
 \begin{equation}\label{eq: main pre} 
\frac{1}{n}\sum_{i=1}^{n}h(q,X^{\top}_{i} \hat{\beta}) +\sum_{\{i:\delta_{i}=1\}} w_{i}\left\{I\left[Y_{i}\leq q\right]-h(q,X_{i} \hat{\beta})\right\} = \tau,
\end{equation}
where $w_i$ for $1\le i \le n$ are data-dependent weights.
The exact solution of  \eqref{eq: main pre} may not exist due to the non-smoothness of the estimating function. However, just as discussed in \cite{han2019quantile} and \cite{Zhang2011Quantile}, any value of $q$ that minimizes the absolute value of the difference between two sides of \eqref{eq: main pre}  can be taken as our estimate and the practical impact of this arbitrariness are negligible in large samples.
 With a given $w=
(w_1, w_2,\cdots, w_n)$, an estimator for $q_{0}$ can be obtained by solving \eqref{eq: main pre}, which is denoted by $\hat{q}_{w}$. 
If $w_i$ for $1\le i\le n$ are taken to be zero, we have $\hat{q}_{w}=\tilde{q}$, which is not $\sqrt{n}$-consistent.
 If the selection probability function $\pi(x)=E\left[\delta\mid X=x\right]$ is specified by a parametric model $g(x^{\top}\gamma)$ and then $w_i$ is taken as  the inverse of $g(X^{\top}_i\hat{\gamma})$ with $\hat{\gamma}$ obtained by the lasso method, the resulting  estimator is the augmented inverse probability weighted estimator in \cite{Belloni2017Quantile}. 
When only one of $f_{Y\mid X}(Y\mid X=x)$ and $\pi(x)$ is correctly specified, the augmented inverse probability weighted estimator may not be $\sqrt{n}$-consistent to $q_0$ due to the same reason as $\tilde{q}$, where $\hat{\beta}$ and $\hat{\gamma}$ are generally not $\sqrt{n}$-consistent due to high dimension of $X$.
This is different from the classical case where the dimension of $X$ is a constant.
However, its asymptotic normality   is proved  when both $f_{Y\mid X}(Y\mid X=x)$ and $\pi(x)$ are correctly specified. 
Clearly, the weights play a crucial role for the asymptotic property of the adjusted estimating equation estimator defined by (2.3).    
Hence, we propose  a debiased method by constructing optimal weights, which are obtained  by solving a convex programming, such that not only the resulting estimator is  asymptotically normal but also its asymptotic variance attains minimum.
This method hence may define an asymptotically more efficient estimator compared to the existing ones and avoids the requirement to specify $\pi(x)$.

This method consists of the following steps:
 \begin{enumerate}[Step 1]
\item Calculate pilot estimators $\hat{\beta}$ and $\tilde{q}$ by solving \eqref{lasso beta} and \eqref{eq: imputation}, respectively.
\item With pilot estimators $\hat{\beta}$ and $\tilde{q}$, construct the debiasing weight $\hat{w}$ as follows
		\begin{alignat}{2}\label{Origin : optim}
		\hat{w} = \argmin_{w} \quad &\sum _{\{i:\delta_i=1\}} w^{2}_{i}h(\tilde{q},X^{\top}_{i}\hat{\beta})(1-h(\tilde{q},X^{\top}_{i}\hat{\beta})) &{}  \\
		\mbox{s.t.}\quad 
		& \left\|\frac{1}{n}\sum_{i=1}^{n}\dot{h}_{u}(\tilde{q},X^{\top}_{i}\hat{\beta})X_{i}-\sum_{\{i:\delta_{i}=1\}}w_{i}\dot{h}_{u}(\tilde{q},X^{\top}_{i}\hat{\beta})X_{i}\right\|_{\infty}\leq\Delta, \label{Origin : constrain beta}\\
		&\sum_{\{i:\delta_{i}=1\}} {w}_{i} = 1\label{Origin : constrain},
		\end{alignat}
		where $\Delta$ is a suitable tuning parameter tending to zero.
\item Replace $w$ in \eqref{eq: main pre} by $\hat{w}$ and solve the following modified equation
 \begin{equation}\label{eq: main prop} 
\frac{1}{n}\sum_{i=1}^{n}h(q,X^{\top}_{i} \hat{\beta}) +\sum_{\{i:\delta_{i}=1\}} \hat{w}_{i}\left\{I\left[Y_{i}\leq q\right]-h(q,X_{i} \hat{\beta})\right\} = \tau.
\end{equation}
\end{enumerate}
The proposed estimator $\hat{q}$ is the solution of \eqref{eq: main prop}. The optimization of the convex programming in Step 2 is to obtain weights such that $\hat{q}$ is root-$n$ consistent, and its asymptotic variance attains minimum which is discussed in the following section.  
   
\section{Asymptotic Properties}\label{sec: 3}
In order to investigate asymptotic properties of the proposed estimator, we first list the following conditions.    
    \begin{itemize}
 \item[(A.1)] (i) The parameter spaces $\mathcal{Q}$ for $q$ is compact, and $q_{0}$ is in the interior of $\mathcal{Q}$. 
 
  (ii) $Y$ is a real value continuous random variable with strictly increasing cumulative distribution funtion.
 \end{itemize}
   \begin{itemize}
\item[(A.2)] $\inf_{x}\pi(x)>0$.
\end{itemize}
 \begin{itemize} 
 \item[(A.3)] (i) $0< \inf_{x}|h(q_0,x^{\top}\beta^0)|\leq \sup_{x}|h(q_0,x^{\top}\beta^0)|<1$.
 
(ii) $\sup_{(q,u)}|\dot{h}_u(q,u)|<\infty$, where $\dot{h}_u(q,u)=\partial h(q,u)/\partial u$. 

(iii)  There exists positive constants $L_{1}$ and $L_2$ such that $|\dot{h}_u(q_1,u_1)-\dot{h}_u(q_{2},u_2)|\leq L_{1}\left(|q_1-q_2|+|u_1-u_2|\right)$ and $|f(q_1, u_1)-f(q_{2},u_2)|\leq L_{2}\left(|q_1-q_2|+|u_1-u_2|\right)$. 
\end{itemize}
 \begin{itemize}
  \item[(A.4)] (i) $\max_{1\leq j\leq p}|X_{j}|\leq K$ almost surely, where $K$ is a positive constant. 
  
  (ii) $\Sigma \equiv E[XX^{\top}]\in\mathbb{R}^{p\times p}$ satisfies that $\Lambda_{\max}\leq c_{\max}<\infty$ and $\Lambda_{\min}\geq c_{\min}>0$, where $\Lambda_{\min}$ and $\Lambda_{\max}$ are the smallest and the largest eigenvalues of $\Sigma$ respectively, and $c_{\max}$ and $c_{\min}$ are positive constants. 
  \end{itemize}  
\begin{itemize}
\item[(A.5)] Let $D = (X,Y,\delta)$ and $\rho_{\beta}$$(x,\tilde{y})$$=-\delta$$\log$$\left(f(y,x^{\top}\beta)\right)$ with $\tilde{y} = (y,\delta)$ be a loss function which is assumed to be a convex function in $\beta$.  (i) There exists positive constants $c$ and $c_m$ such that $E[\rho_{\beta}(D)-\rho_{\beta^{0}}(D)]\geq c_m\|\beta-\beta^{0}\|^2_2$ holds for all $\beta$ with $\|\beta-\beta^0\|_1\leq c$.   
 (ii) For all $\beta$ and $\tilde{\beta}$, $|\rho_{\beta}(x,\tilde{y})-\rho_{\tilde{\beta}}(x,\tilde{y})|\leq L_{\rho}|x^{\top}\beta-x^{\top}\tilde{\beta}|$, where $L_{\rho}$ is a positive constant not depending on $\tilde{y}$. 
\end{itemize}
\begin{remark}
Condition (A.1) is often assumed for quantile estimation, and (A.1)(ii) ensures the identifiability of $q_{0}$. See, for example, \cite{Firpo2007Quantile}, \cite{han2019quantile}. 
Condition (A.2) is fundamental in the missing problem, which means that each individual with the covariates values has positive probability to be observed. 
Condition (A.3) puts some requirements on the data-generating model. 
(A.3)(i) assumes that $P(Y\leq q_{0}\mid X=x)$ is bounded away from zero and one. It is reasonable since $q_0 $ is the $\tau$-quantile of $Y$ with $0<\tau<1$.  (A.3)(ii) is a boundness assumption on the partial derivative of $h(q,u)$ and (A.3)(iii) assumes $h(q,u)$ and $f(q,u)$ satisfy Lipschitz  conditions.
 Condition (A.4)(i) is a commonly used condition in literature of high dimensions. See, for example, Assumption A in \cite{Buhlmann2008high}, (C3) in \cite{Buhlmann2014high}, etc. 
 Condition (A.5) (i) and (ii) are commonly used conditions for the consistency of lasso estimators. See, for example, Assumption L, B in \cite{Buhlmann2008high}, margin condition of Theorem 6.4 and conditions of Theorem 14.5 in \cite{Buhlmann2011uniform}, (A.3) and (A.4) in \cite{Buhlmann2014high}.
 Condition (A.5) (i) holds, for example, when $\rho_{\beta}(x, \tilde{y})$ is twice differentiable on $\beta$, and the expectation  of the second derivative is larger than some positive constant. 
 \end{remark}  

  Define $A = (\dot{h}_u(q_0,X^{\top}\beta^0)X^{\top},1)^{\top}\in \mathbb{R}^{(p+1)\times 1}$ and
\begin{alignat}{2}\label{dual : expe star}
 \eta^*=\argmin_{\eta\in \mathbb{R}^{(p+1)}} \quad &\frac{1}{4}E\left[\frac{\eta^{\top}\delta AA^{\top}\eta}{h(q_{0},X^{\top}\beta^0)(1-h(q_{0},X^{\top}\beta^0))}\right]-E[A^{\top}]\eta. &{} 
\end{alignat}
The following theorem states that the proposed method defines an asymptotically normal estimator of $q_0$. 
\begin{theorem}\label{theo: main}
 Assume $s^4\log(p)=o(\sqrt{n})$ where $s=\|\beta^0\|_0\vee\|\eta^*\|_0$ and the smallest eigenvalue of $E[AA^{\top}]$ is bounded away from zero. Under conditions (A.1)--(A.5), if $\lambda\asymp\sqrt{\log(p)/n}$ and $\Delta\asymp n^{-5/16}(\log(p))^{1/8}$,
  then we have 
\begin{equation}\label{main results}
\sqrt{n}\sigma^{-1}(\hat{q}-q_{0})\stackrel{d}{\rightarrow}N(0,1),
\end{equation}
where $\sigma^2=T^{-2}V$, $T = E[f(q_{0},X^{\top}_{i}\beta^0)]$ and 
$$V=\frac{1}{4}E\left[\frac{\eta^{*\top}\delta AA^{\top}\eta^*}{h(q_{0},X^{\top}\beta^0)(1-h(q_{0},X^{\top}\beta^0))}\right]+\Var\left(h(q_{0},X^{\top}\beta^0)\right).$$
\end{theorem}
 The sparsity condition $s^4\log(p)=o(\sqrt{n})$ seems somewhat stronger than that used by \cite{Belloni2017Quantile}. However, 
\cite{Belloni2017Quantile} need to specify both $f_{Y\mid X}(Y\mid X=x)$ and $\pi(x)$ 
correctly to prove the asymptotic normality 
 of the augmented inverse probability
weighted estimator, and  the proposed method 
defines  an asymptotically normal estimator without specifying $\pi(x)$.  
In addition, the proposed method can define a more efficient estimator than that of \cite{Belloni2017Quantile}. Next, we show this by analyzing $V$ and comparing it with the asymptotic variance in \cite{Belloni2017Quantile}. For simplicity, denote $\tau(X) = h(q_{0},X^{\top}\beta^0)(1-h(q_{0},X^{\top}\beta^0))$.

Note that $\eta^*$ can be explicitly written as follows
 \begin{equation*}
\begin{split}
\eta^*
=2\left(E\left[\frac{\delta AA^{\top}}{\tau(X)}\right]\right)^{-1}E[A].
\end{split}
\end{equation*}
 Then we have
 \begin{equation*}
\begin{split}
V=E[A^{\top}]\left(E\left[\frac{\delta AA^{\top}}{\tau(X)}\right]\right)^{-1}E[A]+\Var\left(h(q_{0},X^{\top}\beta^0)\right).
\end{split}
\end{equation*}
Let $e(X) = \tau(X)^{-1}A$ and
 \begin{equation}\label{eq: r*}
\begin{split}
r^*(X)
=e(X)^{\top}\left(E\left[\delta \tau(X) e(X)e(X)^{\top}\right]\right)^{-1}E[\delta \tau(X)e(X)\pi(X)^{-1}].
\end{split}
\end{equation}
Then we have
 \begin{equation*}
\begin{split}
V=E\left[\delta \tau(X)r^*(X)^2\right]+\Var\left(h(q_{0},X^{\top}\beta^0)\right).
\end{split}
\end{equation*}
The asymptotic variance derived by \cite{Belloni2017Quantile} is $\sigma^2_b=T^{-2}V_b$ where
 \begin{equation*}
\begin{split}
V_b
=&E\left[\delta \tau(X)r(X)^2\right]+\Var\left(h(q_{0},X^{\top}\beta^0)\right)
\end{split}
\end{equation*}
and $r(X) = \pi(X)^{-1}$.
Let $\mathcal{F} = \{f:E[\delta\tau(X)f(X)^{2}]\leq \infty\}$ and define the inner product on $\mathcal{F}$ by $\langle f_{1}, f_{2}\rangle_{\#} = E[\delta \tau(X) f_{1}(X)f_{2}(X)]$. Then $\mathcal{F}$ is a Hilbert space with respect to $\langle \cdot, \cdot\rangle_{\#}$ and the norm induced by the inner product satisfies $\|f\|_{\#}^{2} = E[\delta\tau(X) f(X)^{2}]$. Then we have
 \begin{equation*}
\begin{split}
V=\|r^{*}\|^2_{\#}+\Var\left(h(q_{0},X^{\top}\beta^0)\right).
\end{split}
\end{equation*}
and
 \begin{equation*}
\begin{split}
V_b
=&\|r\|^2_{\#}+\Var\left(h(q_{0},X^{\top}\beta^0)\right).
\end{split}
\end{equation*}
The form of $r^{*}(X)$ in \eqref{eq: r*} indicates that $r^*(X)$ is the projection of $r(X)$ on the space spanned by $e(X)$. Hence we have $\|r^*\|^2_{\#}\leq \|r\|^2_{\#}$ and $\sigma^2\leq \sigma_{b}^2$. The inequality holds if $r(X)$ is not in the space spanned by $e(X)$. 
Note that $V_{b}$ is actually the semiparametric efficiency bound established in \citep{Firpo2007Quantile}. 
Compared to \cite{Firpo2007Quantile}, we make an extra parametric assumption on $f_{Y\mid X}(y\mid X=x)$, our results indicate that this parametric assumption may induce a smaller efficiency bound.

The asymptotic variance of the proposed estimator can be consistently estimated as follows. Define $\hat{\sigma}^2 = (\hat{T})^{-2}\hat{V}$, where $\hat{T} = \frac{1}{n}\sum_{i=1}^{n}f(\tilde{q},X^{\top}_{i}\hat{\beta})$ and $\hat{V}=\hat{V}_1+\hat{V}_2$ with
$$ \hat{V}_1 = n\sum_{\{i:\delta_{i}=1\}}\hat{w}^2_ih(\tilde{q},X^{\top}_i\hat{\beta})\left\{1-h(\tilde{q},X^{\top}_i\hat{\beta})\right\}$$ 
and
 $$ \hat{V}_2 =  \frac{1}{n}\sum_{i=1}^{n}\left[h(\tilde{q},X^{\top}_i\hat{\beta})^2-\left\{\frac{1}{n}\sum_{i=1}^{n}h(\tilde{q},X^{\top}_i\hat{\beta})\right\}^2\right].$$ 

 \begin{theorem}\label{theo: main var}
Under conditions of \Cref{theo: main}, we have 
 $\hat{\sigma}^2\stackrel{p}{\rightarrow}\sigma^2$.
\end{theorem}

The consistent variance estimation ${\hat\sigma}^2$ depends on $\hat{w}$ by  $\hat{V}_1$ only
and $\hat{V}_1$ is just the objective function in \eqref{Origin : optim}--\eqref{Origin : constrain} with $w_i$ replaced by ${\hat w}_i$ for $i=1,2,...,n$.
This makes that the variance estimation and hence the asymptotic variance attains the minimum. The constraint in \eqref{Origin : constrain beta} controls the bias of the resulting 
estimator. 

 \section{Derivation of the optimal weights}\label{sec: 4}
After establishing the theoretical properties of the proposed estimator, we next focus on the computation of the weight $\hat{w}$.
 From Step $1$ and Step 2 of the proposed method, we need to make concrete choice for the tuning parameter $\lambda$ and $\Delta$. 
 We apply the $10$-fold cross validation (CV) method to choose $\lambda$.
  For the selection of tuning parameter $\Delta$, since the asymptotic normality of $\hat{q}$ implies $E[(\hat{q}-q_0)^2]=\sigma^2/n+r(\Delta)$ when $\hat{q}$ is uniform square integrate, where  $r(\Delta)=o(n^{-1})$ is the second-order term of mean square error of $\hat{q}$. Hence  $\Delta$ affects the second-order term of the mean square error of the estimator and hence its selection might not be so critical. 
 According to \Cref{theo: main}, we take $\Delta=cn^{-5/16}(\log(p))^{1/8}$, where c is a positive constant. 
We set $c$ to be $0.10$ which has led to good finite sample performance in our simulations, and if the optimizing problem is infeasible, update $c$ by $0.11,0.12,0.13,\cdots$ in turn until the constraints \eqref{Origin : constrain beta} and \eqref{Origin : constrain} have feasible points. 
  Furthermore, to ensure the numerical implementation, we provide an equivalent easy-to-implement alternative to compute $\hat{w}$.
According to \cite{Athey2018debias}, we can establish a $1:1$ mapping between the optimizing problem \eqref{Origin : optim}--\eqref{Origin : constrain} and the following optimizing problem
\begin{equation}\label{Origin : optim 2}
\begin{split}
\hat{w} =& \argmin_{w} \left[(1-\hat{\zeta})\sum _{\{i:\delta_i=1\}} w^{2}_{i}h(\tilde{q},X^{\top}_{i}\hat{\beta})(1-h(\tilde{q},X^{\top}_{i}\hat{\beta}))\right.\\
&\left.+\hat{\zeta} \left\|\frac{1}{n}\sum_{i=1}^{n}\dot{h}_{u}(\tilde{q},X^{\top}_{i}\hat{\beta})X_{i}-\sum_{\{i:\delta_{i}=1\}}w_{i}\dot{h}_{u}(\tilde{q},X^{\top}_{i}\hat{\beta})X_{i}\right\|^2_{\infty}\right] \\
&s.t.\quad  \begin{matrix}
&\sum_{\{i:\delta_{i}=1\}} {w}_{i} = 1,
\end{matrix}
\end{split}
\end{equation}
where 
\begin{equation}\label{Origin : optim 2 zeta}
\begin{aligned}
\hat{\zeta} 
=& \argmax_{\zeta\in [0, 1)}\min_{w} \left[\sum _{\{i:\delta_i=1\}} w^{2}_{i}h(\tilde{q},X^{\top}_{i}\hat{\beta})(1-h(\tilde{q},X^{\top}_{i}\hat{\beta}))\right.\\
		&\left.+\frac{\zeta}{(1-\zeta)} \left\|\frac{1}{n}\sum_{i=1}^{n}\dot{h}_{u}(\tilde{q},X^{\top}_{i}\hat{\beta})X_{i}-\sum_{\{i:\delta_{i}=1\}}w_{i}\dot{h}_{u}(\tilde{q},X^{\top}_{i}\hat{\beta})X_{i}\right\|^2_{\infty}-\frac{\zeta}{(1-\zeta)}\Delta^2 \right]\\
&\begin{array}{r@{\quad}r@{}l@{\quad}l}
s.t. &\sum_{\{i:\delta_{i}=1\}} {w}_{i} = 1
\end{array}
\end{aligned}
\end{equation}
is the solution to the dual problem of \eqref{Origin : optim}--\eqref{Origin : constrain}.
Note that \eqref{Origin : optim 2} is equal to 
\begin{equation}\label{Origin : optim 3}
\begin{split}
\hat{w} =& \argmin_{w}\min_{\Gamma}\left[(1-\hat{\zeta})\sum _{\{i:\delta_i=1\}} w^{2}_{i}h(\tilde{q},X^{\top}_{i}\hat{\beta})(1-h(\tilde{q},X^{\top}_{i}\hat{\beta}))+\hat{\zeta} \Gamma^2\right] \\
&s.t.\quad  \begin{matrix}
& \left\|\frac{1}{n}\sum_{i=1}^{n}\dot{h}_{u}(\tilde{q},X^{\top}_{i}\hat{\beta})X_{i}-\sum_{\{i:\delta_{i}=1\}}w_{i}\dot{h}_{u}(\tilde{q},X^{\top}_{i}\hat{\beta})X_{i}\right\|_{\infty}\leq \Gamma,\\
&\sum_{\{i:\delta_{i}=1\}} {w}_{i} = 1,
\end{matrix}
\end{split}
\end{equation}
and the optimizing problem \eqref{Origin : optim 3} is a quadratic programming. 
Then we can obtain $\hat{w}$ by using the \texttt{solve.QP} from the \texttt{quadprog} package.

 The calculation of $\hat{\zeta}$ can be achieved  by the following three steps. 
 \begin{enumerate}[Step 1]
\item Consider a set of values $G=\{0,0.01,\cdots,0.98,0.99\}$  for $\zeta$ and denote the $l$-th value by $\zeta_l$.
\item For each $\zeta_l$ in $G$, calculate a weight by \eqref{Origin : optim 3} with $\hat{\zeta}$ replaced by $\zeta_l$, and denote the weight by $\hat{w}(\zeta_l)$.
\item With $\hat{w}(\zeta_l)$ for all $\zeta_l\in G$, according to \eqref{Origin : optim 2 zeta}, approximate $\hat{\zeta}$ by
\begin{align*}
\hat{\zeta}
=& \argmax_{\zeta_l\in G} \left[\sum _{\{i:\delta_i=1\}} \hat{w}(\zeta_l)^{2}_{i}h(\tilde{q},X^{\top}_{i}\hat{\beta})(1-h(\tilde{q},X^{\top}_{i}\hat{\beta}))\right.\\
		&\left.+\frac{\zeta_l}{(1-\zeta_l)} \left\|\frac{1}{n}\sum_{i=1}^{n}\dot{h}_{u}(\tilde{q},X^{\top}_{i}\hat{\beta})X_{i}-\sum_{\{i:\delta_{i}=1\}}\hat{w}(\zeta_l)_{i}\dot{h}_{u}(\tilde{q},X^{\top}_{i}\hat{\beta})X_{i}\right\|^2_{\infty}\right.\\
		&\left.-\frac{\zeta_l}{(1-\zeta_l)}\Delta^2 \right].
\end{align*}
\end{enumerate} 
Then according to \eqref{Origin : optim 2}, the solution to the optimizing problem \eqref{Origin : optim}--\eqref{Origin : constrain} is given by $\hat{w}=\hat{w}(\hat{\zeta})$. 
In contrast to \cite{Athey2018debias}, which simply takes $\hat{\zeta}=0.5$ for every $\Delta$, we provide a data-adaptive and reasonable computing way to obtain $\hat{\zeta}$.

\section{Simulation Study}\label{sec: 5}
To evaluate the numerical performance of the proposed method, we conducted a simulation study with the design similar to \cite{Tan2020high} and calculated the augmented inverse probability weighting (AIPW) estimator $\hat{q}_{aipw}$ due to \cite{Belloni2017Quantile} as a comparison. 
Unless otherwise specified, a logistic model of $\delta$ versus $X$ and a standard normal linear conditional distribution of $Y$ given $X$ were assumed for $\pi(X)$ and $f_{Y\mid X}(y\mid X)$, respectively.

Let $X = (X_1,\cdots,X_p)$, where $X_{j}$ for $j=1$ and $2$ was generated from a uniform distribution $U(-5,5)$ and $X_j$ for $j=3,\cdots,p$ from a truncated normal with mean $0$, variance $1/2$ and truncation constant $5$. 
In addition, let $X^{\dagger} = (X^{\dagger} _1,\cdots,X^{\dagger} _p)$, where $X^{\dagger} _j=X_j-X^2_j+2X^3_j$ for $j=1,2,3$ and $4$ and $X^{\dagger} _j=X_j$ for $j=5,\cdots,p$. 
Consider the median case ($\tau=0.5$) under the following two data-generating processes (DGP): 
\begin{enumerate}[(DGP1)]
\item Generate $Y$ given $X$ from a normal distribution
\[N(0.25X_1+0.125X_2+0.25X_3+0.125X_4,1)\] 
and generate $\delta$ given $X$ from a Bernoulli distribution with 
\[P(\delta=1\mid X)=\frac{\exp(1-0.25X^{\dagger}_1-0.125X^{\dagger}_2-0.25X^{\dagger}_3-0.125X^{\dagger}_4)}{1+\exp(1-0.25X^{\dagger}_1-0.125X^{\dagger}_2-0.25X^{\dagger}_3-0.125X^{\dagger}_4)}.\]
\item Generate $Y$ given $X$ as in DGP1 but generate $\delta$ given $X$ from a Bernoulli distribution with 
\[P(\delta=1\mid X)=\frac{\exp(1-0.25X_1-0.125X_2-0.25X_3-0.125X_4)}{1+\exp(1-0.25X_1-0.125X_2-0.25X_3-0.125X_4)}.\]
\end{enumerate} 
  Depending on above data generation processes where $\pi(X)$  involves two completely different sets of regressors, $\pi(X)$ is misspecified under DGP1 and correctly specified under DGP2. 

For each of the two DGPs, the simulation was conducted based on $1000$ replications with sample size of $n = $ $200$, $400$ and $800$ and covariates number of $p=$$\frac{1}{4}n$, $\frac{1}{2}n$, $n$ and $2n$, respectively. 
From the $1000$ simulated values of $\hat{q}_{aipw}$ and $\hat{q}$, we computed the Monte Carlo bias (Bias), standard deviation (SD), root mean square error (RMSE). For nominal confidence level $1-\alpha=0.95$, we evaluated the coverage probabilities (CP) of the confidence intervals.  
The simulation results are reported in \Cref{tab 1-1} and \Cref{tab 1-2} for DGP1 and DGP2, respectively.  
In addition, we compared the estimated standard deviation of $\hat{q}$ (ESD) based on the asymptotic variance given by \cref{theo: main var} with the Monte Carlo standard deviation based on $1000$ repetitions, which are reported in \Cref{tab 2}.

 \centerline{[Insert \Cref{tab 1-1}, \Cref{tab 1-2} and \Cref{tab 2} about here.]}

From \Cref{tab 1-1}, \Cref{tab 1-2} and \Cref{tab 2}, we have the following observations.
\begin{enumerate}[(i)]
\item  In the case where $\pi(X)$ is misspecified, $\hat{q}$ outperforms $\hat{q}_{aipw}$ in terms of Bias, RMSE and CP for all combinations of $n$ and $p$, especially when $p$ diverges with $n$ at a relatively large rate. 
Although $\hat{q}_{aipw}$ has generally slightly smaller SD than $\hat{q}$, its Bias is approximately $5$ times as large as that of $\hat{q}$. 
In addition, the coverage probability based on of the AIPW estimator is considerably lower than the nominal level $95\%$. 
On the contrary,  the proposed estimator performs well with coverage probabilities generally closing to $0.95$, which is expected.
This could be explained by the fact that the root-$n$ asymptotic normality of $\hat{q}_{aipw}$ requires that $\pi(X)$ is correctly specified and the requirement is not satisfied in this case. The fact also implies that the AIPW method cannot be used to make statistical inference for $q_0$ in the case where $\pi(X)$ is misspecified. 
On the contrary, the calculation of the proposed estimator $\hat{q}$ does not involve $\pi(X)$ and hence the asymptotic normality of $\hat{q}$ is robust to the misspecification of $\pi(X)$.

\item In the case where $\pi(X)$ is correctly specified, both $\hat{q}$ and $\hat{q}_{aipw}$ perform well while the standard deviations of $\hat{q}$ are generally smaller than those of $\hat{q}_{aipw}$, which is in agreement with the asymptotics in \cref{theo: main var}. 
\item The estimated standard deviations are close to the empirical standard deviations for the proposed estimator $\hat{q}$.
\end{enumerate}
Overall, our theoretical results are supported by the simulation studies. Although $\hat{q}_{aipw}$ has comparable performance to $\hat{q}$ in the case where $\pi(X)$ is correctly specified, it is hard to specify a correct model for $\pi(X)$ in practice. Hence the proposed method is more trustworthy and hence recommended.  

In addition, our simulation results indicate that the proposed estimator performs fairly well even if this sparsity condition $s=o(n^{1/8}(\log(p))^{-1/4})$ is violated. This implies that the sparsity condition may be weaken.  In Section S3 of the supplementary material, we weaken the sparsity condition while maintaining the $\sqrt{n}$ consistency via data splitting

\section{Real Data Analysis}\label{sec: 6}
We provide an application to analyzing a medical dataset collected on 2139 HIV-infected subjects enrolled in AIDS clinical Trial Group Protocol 175 (ACTG 175). 
 The original data were collected by \cite{hammer1996actg}. ACTG 175 is a randomized clinical trial where patients are randomized to four antiretroviral regimens in: zidovudine (ZDV) only, ZDV+didanosine (ddI), ZDV+zalcitabine (ddC), and ddI only. 
 Following the analysis in \cite{Davidian2005actg}, we consider two groups: the group with ZDV alone (control) and the group with the other three therapies (treatment). 
The dataset contains $n=2139$ patients,  $n_0 = 532$ of whom participated in the control group and $n_1=1607$ of whom participated in the treated group ($treat$: 0=control).  
 This study evaluates the treatment effect by the change in CD4 count from baseline to $96\pm 5$ weeks (CD4$_{96}$) which is a measure of immunologic status. 
 Previous work analyzed the dataset by the average treatment effect (see, \cite{Davidian2005actg}, \cite{han2014regression} and \cite{han2019quantile}). 
 Our main interest is the median treatment effect $m = m_1-m_0$, where $m_1$ and $m_0$ are the median of CD4$_{96}\mid treat = c$ with $c=1$ and $0$ respectively.
 
   However, there are $797$ subjects whose CD4$_{96}$ are missing ($r$: 0=missing) due to dropout from the study. At the baseline and during the follow-up, $23$ covariates ($X$) correlated with CD4$_{96}$ are obtained. 
   There may be interactions between covariates $X$.  To employ the proposed method, we treat  the observed covariates and their two-way interactions as a new covariate vector. Specifically, we denote it by a vector $U$  whose $j$-th component $U_j=X_j$ for $1\leq j\leq 23$, and $U_j=X_lX_{j-23l+l(l-1)/2}$ for $[23l-(l-1)(l-2)/2]< j\leq [23(l+1)-l(l-1)/2]$ and $l=1,2,\cdots,23$. The dimension of $U$ is $299$.    
As analyzed in \Cref{sec: 3}, the selection probability function is unknown and difficult to specify correctly, in which case AIPW method performs poorly and cannot be used to make inference, and hence  we apply  our method to the real data analysis only here.   
 We consider a standard normal distribution for $CD4_{96}\mid U, treat=c$ with $c=0,1$. The confidence intervals for $m_1$, $m_0$ and $m$ obtained via the proposed method are reported in \Cref{tab 3}.
 
 \centerline{[Insert \Cref{tab 3} about here.]}
 
 From \Cref{tab 3}, it can be seen that people who received three newer treatments had a potential median effect of $\hat{m} = 48$ compared with those who received ZDV alone. Moreover, this effect is significant since the $95\%$ confidence interval does not contain $0$.

\noindent {\large\bf Supplementary Materials}
Supplementary materials are available online, which contain the lemmas that used in the proofs of \Cref{theo: main} and \Cref{theo: main var}, some additional simulation results for \Cref{sec: 5} and further explorations about the sparsity condition. \par

\par
\section*{Appendix. Proof of Main Results}
Appendix contains proofs of \Cref{theo: main} and \Cref{theo: main var}. Note that constant $c$ may vary from lines and all of them are positive.
\begin{proof}[Proof of \Cref{theo: main}]
For the simplicity of illustration, let
 \[\hat{F}_n\left(q\right)\triangleq\frac{1}{n}\sum^n_{i=1}h(q,X^{\top}_{i} \hat{\beta}) +\sum_{\{i:\delta_{i}=1\}}\hat{w}_{i}\hat{\xi}_{i}\left(q\right) ,\]
where $\hat{\xi}_{i}\left(q\right) = I\left[Y_{i}\leq q\right]-h(q,X^{\top}_{i}\hat{\beta})$. 
 An outline of the proof of \Cref{theo: main} is as follows. According to mean value theorem, it follows that
\begin{equation}\label{norm: decom}
\begin{split}
\hat{F}_n\left(\hat{q}\right) - \hat{F}_n\left(q_{0}\right)  
=& \frac{1}{n}\sum_{i=1}^{n}f(\bar{q},X^{\top}_{i}\hat{\beta})\left(\hat{q}-q_{0}\right) + \sum_{\{i:\delta_{i}=1\}}\hat{w}_{i}\left\{\hat{\xi}_{i}\left(\hat{q}\right) - \hat{\xi}_{i}\left(q_{0}\right)\right\},
\end{split}
\end{equation}
where $\bar{q}$ is between $q_{0}$ and $\hat{q}$.  
On the one hand, we show that 
  \begin{equation}\label{taylor 1}
  \frac{1}{n}\sum_{i=1}^{n}f(\bar{q},X^{\top}_{i}\hat{\beta})\left(\hat{q}-q_{0}\right) =\left\{E\left[f(q_{0},X^{\top}\beta^0)\right]+o_p(1)\right\}(\hat{q}-q_{0})
  \end{equation}
  and
 \begin{equation}\label{taylor 2}
\sum_{\{i:\delta_{i}=1\}}\hat{w}_{i}\left\{\hat{\xi}_{i}\left(\hat{q}\right) - \hat{\xi}_{i}\left(q_{0}\right)\right\}=o_p\left(|\hat{q}-q_{0}|\right)+o_p\left( n^{-1/2}\right).
\end{equation}
  On the other hand, we show that
\begin{equation}\label{taylor}
 \begin{split}
\hat{F}_n\left(\hat{q}\right) - \hat{F}_n\left(q_{0}\right)  
=& -\frac{1}{n}\sum_{\{i:\delta_{i}=1\}}\frac{A^{\top}_{i}\eta^*\xi_{i}\left(q_{0}\right)}{2h(q_{0},X^{\top}_i\beta^0)(1-h(q_{0},X^{\top}_i\beta^0))}\\
&-\left\{\frac{1}{n} \sum_{i=1}^nh(q_{0},X^{\top}_i\beta^0)-E\left[h(q_0,X^{\top}\beta^0)\right]\right\}\\
&+o_p\left(n^{-1/2}\right),
\end{split}
\end{equation}
where $A_i = (\dot{h}_u(q_0,X^{\top}_i\beta^0)X^{\top}_i,1)^{\top}\in \mathbb{R}^{(p+1)\times 1}$, $\xi_i(q)=I[Y_i\leq q]-h(q,X^{\top}_i\beta^0)$ and $\eta^*$ is defined in \eqref{dual : expe star}. \eqref{taylor 1}--\eqref{taylor} together with \eqref{norm: decom} implies the following asymptotic representation
\begin{equation}\label{ALP}
\begin{split}
&-\left\{E[f(q_{0},X^{\top}\beta^0)]+o_p(1)\right\}\left(\hat{q}-q_{0}\right)\\
=& \frac{1}{n}\sum_{\{i:\delta_{i}=1\}}\frac{A^{\top}_{i}\eta^*\xi_{i}\left(q_{0}\right)}{2h(q_{0},X^{\top}_i\beta^0)(1-h(q_{0},X^{\top}_i\beta^0))}\\
&+\left\{\frac{1}{n} \sum_{i=1}^nh(q_{0},X^{\top}_i\beta^0)-E\left[h(q_0,X^{\top}\beta^0)\right]\right\}+o_p\left(n^{-1/2}\right).
\end{split}
\end{equation}
Then the main result \eqref{main results} in \Cref{theo: main} is proved by the central limit theorem and Slutsky's theorem.

 (a) First, we prove \eqref{taylor 1}. Lemma 1 in the supplementary material proves that 
\begin{equation}\label{eq: beta 1}
\left\|\hat{\beta}-\beta^0\right\|_1 = O_p\left(\left\|\beta^0\right\|_0\sqrt{\frac{\log(p)}{n}}\right).
\end{equation}
Then we have $\|\hat{\beta}-\beta^0\|_1 =o_p(1)$ by the assumption on $\|\beta^0\|_0$. If the consistency of $\hat{q}$ is proved, \eqref{taylor 1} follows immediately by the law of large numbers. 
To prove the consistency of $\hat{q}$, on the basis of Theorem 5.9 in \cite{van2000asymptotic} and Condition (A.1), it suffices to check that
 \begin{equation}\label{consis: goal}
\sup_{q}\Big|\hat{F}_n\left(q\right) - E\left[h(q,X^{\top}\beta^0)\right]\Big| = o_{p}\left(1\right).
\end{equation}
Let $F_n\left(q\right)\triangleq n^{-1}\sum^n_{i=1}h(q,X^{\top}_i\beta^0)$. Then we can write
 \begin{equation}\label{eq: consis}
 \begin{split}
&\sup_{q}\Big|\hat{F}_n\left(q\right) - E\left[h(q,X^{\top}_1\beta^0)\right]\Big| \\
\leq& \sup_{q}\Big|\hat{F}_n\left(q\right) - F_n\left(q\right)\Big|+\sup_{q}\Big|F_n\left(q\right) - E\left[h(q,X^{\top}\beta^0)\right]\Big|\\
\triangleq&U_{n1}+U_{n2}.
\end{split}
\end{equation}
First, $U_{n2}=o_p\left(1\right)$ can be proved according to Lemma 2.4 in \cite{Newey1994Large}. Next, we prove $U_{n1}=o_p(1)$.
 By mean value theorem, we have
 \begin{equation}\label{Un: decom}
 \begin{split}
U_{n1}
 \leq&  \sup_{q}\Big|\frac{1}{n}\sum^n_{i=1}\dot{h}_{u}(q,X^{\top}_{i} \tilde{\beta})X^{\top}_i(\hat{\beta}-\beta^0)\Big|\\
 &+\sup_{q}\Big|\sum_{\{i:\delta_{i}=1\}}\hat{w}_{i}\dot{h}_{u}(q,X^{\top}_{i} \bar{\beta})X^{\top}_i(\hat{\beta}-\beta^0)\Big|\\
& +\sup_{q}\Big|\sum_{\{i: \delta_{i}=1\}}\hat{w}_{i}\xi_{i}\left(q\right)\Big|\\
 \triangleq&U_{n11}+U_{n12}+U_{n13},
\end{split}
\end{equation}
where $\tilde{\beta}$ and $\bar{\beta}$ are between $\beta^0$ and $\hat{\beta}$. Then the problem reduces to show that $U_{n1i}=o_p(1)$ for $i=1,2$ and $3$.

By Conditions (A.3)(ii), (A.4)(i) and \eqref{eq: beta 1}, we have 
\[U_{n11}\leq c\left\|\hat{\beta}-\beta^0\right\|_1 = o_p(1)\]
 and 
 \[U_{n12}
\leq c\sum_{\{i: \delta_{i}=1\}}|\hat{w}_{i}|\left\|\hat{\beta}-\beta^0\right\|_1 = o_p\left(\sum_{\{i: \delta_{i}=1\}}|\hat{w}_{i}|\right).\] 
Then $U_{n12}=o_p(1)$ is proved if we can show that $\sum_{\{i:\delta_{i}=1\}}|\hat{w}_{i}|= O_p(1)$. Lemma 3 in the supplementary material proves that the constraints \eqref{Origin : constrain beta} and \eqref{Origin : constrain} can be satisfied with probability tending to $1$ by taking $w_{i}$ to be 
$\tilde{w}_i = \pi\left(X_i\right)^{-1}/\sum_{\{i:\delta_i=1\}}\pi\left(X_i\right)^{-1}$. Recalling the definition of $\hat{w}$, note that $\hat{w}$ not only satisfies the constraints \eqref{Origin : constrain beta} and \eqref{Origin : constrain}, but also minimizes the objective function in \eqref{Origin : optim}. Then we have 
 \begin{equation}\label{eq: comp}
 \begin{split}
 \sum_{\{i:\delta_{i}=1\}}\hat{w}^2_{i}h(\tilde{q},X^{\top}_i\hat{\beta})(1-h(\tilde{q},X^{\top}_i\hat{\beta}))\leq \sum_{\{i:\delta_{i}=1\}}\tilde{w}^2_{i}h(\tilde{q},X^{\top}_i\hat{\beta})(1-h(\tilde{q},X^{\top}_i\hat{\beta})).
 \end{split}
 \end{equation}
By Condition (A.2), the right side of \eqref{eq: comp} is $O_p(n^{-1})$. This together with Condition (A.3)(i) proves $\sum_{\{i:\delta_{i}=1\}}\hat{w}^2_{i}= O_p(n^{-1})$. Since $\left(\sum_{\{i:\delta_{i}=1\}}|\hat{w}_{i}|\right)^2\leq n\sum_{\{i:\delta_{i}=1\}}\hat{w}^2_{i}$ by Cauchy-Schwartz inequality, it follows that $\sum_{\{i:\delta_{i}=1\}}|\hat{w}_{i}|= O_p(1)$.
 
By \eqref{Un: decom}, it remains to show that $U_{n13}=o_p(1)$.  
By the Lagrange multiplier method, Lemma 4 in the supplementary material provides an alternative representation of $\hat{w}_i$ as follows
\[\hat{w}_i = \frac{1}{2n}\frac{\hat{A}^{\top}_i\hat{\eta}}{h(\tilde{q},X^{\top}_i\hat{\beta})(1-h(\tilde{q},X^{\top}_i\hat{\beta}))},\]
 where $\hat{A}_i = (\dot{h}_u(\tilde{q},X^{\top}_i\hat{\beta})X^{\top}_i,1)^{\top}\in \mathbb{R}^{(p+1)\times 1}$ and 
\begin{equation*}\label{dual : expe}
 \hat{\eta}=\argmin_{\eta\in \mathbb{R}^{(p+1)}}  \frac{1}{4n}\sum^n_{i=1}\frac{\eta^{\top}\delta_i\hat{A}_i\hat{A}^{\top}_i\eta}{h(\tilde{q},X^{\top}_i\hat{\beta})(1-h(\tilde{q},X^{\top}_i\hat{\beta}))}-\frac{1}{n}\sum^n_{i=1}\hat{A}^{\top}_i\eta+\left\|\eta_{-(p+1)}\right\|_1\Delta. 
\end{equation*}
Define 
\begin{equation}\label{eq: w star}
w^*_{i}=\frac{1}{2n}\frac{A^{\top}_i\eta^*}{h(q_{0},X^{\top}_i\beta^0)(1-h(q_{0},X^{\top}_i\beta^0))}
\end{equation}
for $i\in\{i:\delta_i=1\}$.
Then we can write
\begin{equation}\label{eq: Un13}
\begin{split}
U_{n13}
\leq&\sup_{q}\Big|\sum_{\{i:\delta_{i}=1\}}(\hat{w}_{i}-w^*_i)\xi_{i}\left(q\right)\Big|+\sup_{q}\Big|\sum_{\{i:\delta_{i}=1\}}w^*_{i}\xi_{i}\left(q\right)\Big|\\
 \triangleq&J_{n1}+J_{n2}.
 \end{split}
\end{equation}
According to the restrictions on the sparsity $s$ and the rate of $\Delta$, the conditions of Lemma 6 in the supplementary material is satisfied.
Then Lemma 6 shows that
\begin{equation}\label{eq: w 1}
\left\|\hat{w}-w^*\right\|_1=o_p(1),
\end{equation}
which implies $J_{n1}=o_p(1)$. By Theorem 37 in \cite{Pollard1984uniform} and Lemma 6 in the supplementary material, we have $J_{n2}=o_p(1)$. These together with \eqref{eq: Un13} prove $U_{n13}=o_p(1)$. Then the consistency of $\hat{q}$ is proved.
 
 (b) Next, we prove \eqref{taylor 2}. It is noted that
\begin{equation}\label{Rn: decom}
\begin{split}
\Big|\sum_{\{i:\delta_{i}=1\}}\hat{w}_{i}\left(\hat{\xi}_{i}\left(\hat{q}\right) - \hat{\xi}_{i}\left(q_{0}\right)\right)\Big|
\leq&\Big|\sum_{\{i:\delta_{i}=1\}}w^*_{i}\left\{\hat{\xi}_{i}\left(\hat{q}\right) - \hat{\xi}_{i}\left(q_{0}\right)\right\}\Big|\\
&+\Big|\sum_{\{i:\delta_{i}=1\}}\left(\hat{w}_{i}-w^*_{i}\right)\left\{\hat{\xi}_{i}\left(\hat{q}\right) - \hat{\xi}_{i}\left(q_{0}\right)\right\}\Big|\\
\triangleq&R_{n1}+R_{n2}.
\end{split}
\end{equation}

 By some algebras and mean value theorem, it follows that
\begin{equation}\label{Rn1: decom}
\begin{split}
|R_{n1}|
 \leq& \Big|\sum_{\{i:\delta_{i}=1\}}w^*_{i}\left\{\xi_{i}\left(\hat{q}\right) - \xi_{i}\left(q_{0}\right)\right\}\Big|\\
&+\Big|\sum_{\{i:\delta_{i}=1\}}w^*_{i}\left\{h(\hat{q},X^{\top}_{i}\hat{\beta})-h(q_{0},X^{\top}_i\hat{\beta})-\left[h(\hat{q},X^{\top}_{i}\beta^0)-h(q_{0},X^{\top}_{i}\beta^0)\right]\right\}\Big|\\
\leq&\sup_{|q-q_{0}| = O(\iota_n)}\Big|\sum_{\{i:\delta_{i}=1\}}w^*_{i}\left\{\xi_{i}\left(q\right) - \xi_{i}\left(q_{0}\right)\right\}\Big|\\
&+\Big|\sum_{\{i:\delta_{i}=1\}}w^{*}_{i}\left[f(\bar{q},X^{\top}_{i}\hat{\beta})-f(\breve{q},X^{\top}_i\beta^{0})\right](\hat{q}-q_0)\Big|\\
\triangleq& R_{n11}+R_{n12},
\end{split}
\end{equation}
where $\bar{q}$ and $\breve{q}$ are between $q_0$ and $\hat{q}$.
Denote $\iota_n$ the convergence rate of $\hat{q}$. According to the Theorem 37 in \cite{Pollard1984uniform} and Lemma 6 in the supplementary material, we have
\begin{equation}\label{Rn11: res}
R_{n11}=o_p(n^{-1/2}\vee \iota_n).
\end{equation}
 By Conditions (A.3)(iii) and (A.4)(i), we have
\begin{align*}
R_{n12}
\leq &c\sum_{\{i:\delta_{i}=1\}}|w^*_i|\left\{|\bar{q}-\breve{q}|+|X^{\top}_i(\hat{\beta}-\beta^0)|\right\}|\hat{q}-q_0|\\
\leq &c\sum_{\{i:\delta_{i}=1\}}|w^*_i|\left\{|\bar{q}-\breve{q}|+\left\|\hat{\beta}-\beta^0\right\|_1\right\}|\hat{q}-q_0|.
\end{align*}
Similar to the previous result,  we can show that $\sum_{\{i:\delta_{i}=1\}}|w^{*}_i|=O_p(1)$. This together with \eqref{eq: beta 1} and the consistency of $\hat{q}$ proves 
\begin{equation}\label{Rn12: res}
R_{n12} = o_p(\iota_n).
\end{equation}
Equations \eqref{Rn1: decom}, \eqref{Rn11: res} and \eqref{Rn12: res} prove
\begin{equation}\label{Rn1: res}
R_{n1} = o_p(n^{-1/2})+o_p(\iota_n).
\end{equation}

By some algebras, it is easy to show that
\begin{equation}\label{Rn2: decom}
\begin{split}
R_{n2}
\leq&\Big|\sum_{\{i:\delta_{i}=1\}}\left(\hat{w}_{i}-w^*_{i}\right)\left\{\xi_i(\hat{q})-\xi_i(q_{0})\right\}\Big|\\
&+\Big|\sum_{\{i:\delta_{i}=1\}}\left\{\hat{w}_{i}-w^*_{i}\right\}\left\{h\left(\hat{q},X^{\top}_{i}\hat{\beta}\right)-h\left(q_{0},X^{\top}_{i}\hat{\beta}\right)\right\}\Big|\\
&+\Big|\sum_{\{i:\delta_{i}=1\}}\left\{\hat{w}_{i}-w^*_{i}\right\}\left\{h\left(\hat{q},X^{\top}_{i}\beta^0\right)-h\left(q_{0},X^{\top}_{i}\beta^0\right)\right\}\Big|\\
\triangleq&R_{n21}+R_{n22}+R_{n23}.
\end{split}
\end{equation}
 By Cauchy-Schwartz inequality, we have
\begin{equation}\label{eq: Rn21}
\begin{split}
R_{n21}
\leq &\left\{n\sum_{\{i:\delta_{i}=1\}}\left(\hat{w}_i-w^*_i\right)^2\right\}^{1/2}\left\{\frac{1}{n}\sum_{i=1}^{n}\left[\xi_i(\hat{q})-\xi_i(q_{0})\right]^2\right\}^{1/2}\\
\leq &\left\{n\sum_{\{i:\delta_{i}=1\}}(\hat{w}_i-w^*_i)^2\right\}^{1/2}\left\{\sup_{|q-q_{0}| = O(\iota_n)}\frac{1}{n}\sum_{i=1}^{n}[\xi_i(q)-\xi_i(q_{0})]^2\right\}^{1/2}.
\end{split}
\end{equation}
Notice that
\begin{equation}\label{Rn21 right}
\begin{split}
&\sup_{|q-q_{0}| = O(\iota_n)}\frac{1}{n}\sum_{i=1}^{n}\left\{\xi_i(q)-\xi_i(q_{0})\right\}^2\\
\leq &\sup_{|q-q_{0}| = O(\iota_n)}\left\{\frac{1}{n}\sum_{i=1}^{n}\left\{\xi_i(q)-\xi_i(q_{0})\right\}^2-E\left[\left\{\xi_i(q)-\xi_i(q_{0})\right\}^2\right]\right\}\\
&+\sup_{|q-q_{0}| = O(\iota_n)}E\left[\left\{\xi_i(q)-\xi_i(q_{0})\right\}^2\right]\\
=&\sup_{|q-q_{0}| = O(\iota_n)}\left\{\frac{1}{n}\sum_{i=1}^{n}\left\{\xi_i(q)-\xi_i(q_{0})\right\}^2-E\left[\left\{\xi_i(q)-\xi_i(q_{0})\right\}^2\right]\right\}+O(\iota_n)\\
=&O_p(n^{-1/2}\vee \iota_n)
\end{split}
\end{equation}
by Theorem 37 in \cite{Pollard1984uniform}. 
In addition, Lemma 6 in the supplementary material proves that 
\begin{equation}\label{eq: w square}
n\left\|\hat{w}-w^*\right\|^2_2=o_p(n^{-1/2}).
\end{equation}
Equations \eqref{eq: Rn21}, \eqref{Rn21 right} and \eqref{eq: w square} imply
\begin{equation}\label{Rn21: res}
R_{n21}=o_p\left(\left(n^{-1/4}\right)\left(n^{-1/4}\vee\iota^{1/2}_n\right)\right)=o_p\left(\left(n^{-1/4}\vee\iota^{1/2}_n\right)^2\right)=o_p(n^{-1/2}\vee\iota_n).
\end{equation}
By mean values theorem and Condition (A.3)(ii), we have 
\begin{align*}
R_{n22}
\leq &c\left\|\hat{w}-w^*\right\|_1|\hat{q}-q_{0}|.
\end{align*}
This together with \eqref{eq: w 1} proves 
\begin{equation}\label{Rn22: res}
R_{n22} =  o_p(|\hat{q}-q_{0}|).
\end{equation}
Similarly, it can be proved that 
\begin{equation}\label{Rn23: res}
R_{n23} =  o_p(|\hat{q}-q_{0}|). 
\end{equation}
Then equations \eqref{Rn2: decom}, \eqref{Rn21: res}, \eqref{Rn22: res} and \eqref{Rn23: res} prove 
\begin{equation}\label{Rn2: res}
R_{n2} = o_p(n^{-1/2})+o_p(\iota_n). 
\end{equation}
Relations \eqref{Rn: decom}, \eqref{Rn1: res} and \eqref{Rn2: res} together prove \eqref{taylor 2}.

(c) Finally, we prove \eqref{taylor}. Note that $\hat{F}_n\left(\hat{q}\right)=\tau = E[h(q_0,X^{\top}\beta^0)]$. Then we have
\begin{equation}\label{eq: ano repre}
\begin{split}
 \hat{F}_n\left(\hat{q}\right) - \hat{F}_n\left(q_{0}\right) 
& = E\left[h(q_{0},X^{\top}\beta^0)\right]- \hat{F}_n\left(q_{0}\right) \\
&= \left\{E\left[h(q_{0},X^{\top}\beta^0)\right]-F_{n}\left(q_{0}\right)\right\}-  \left\{\hat{F}_n\left(q_{0}\right)-F_{n}\left(q_{0}\right)\right\}.
\end{split}
\end{equation}
By mean value theorem and some algebras, it follows that
\begin{equation}\label{Ln decomp}
\begin{split}
&\hat{F}_n\left(q_{0}\right)-F_{n}\left(q_{0}\right)\\
 =& \left[\frac{1}{n}\sum_{i=1}^{n}\dot{h}_{u}(\tilde{q},X^{\top}_{i}\hat{\beta})X^{\top}_i-\sum_{\{i:\delta_{i}=1\}} \hat{w}_{i}\dot{h}_{u}(\tilde{q},X^{\top}_{i}\hat{\beta})X^{\top}_i\right]\left(\hat{\beta}-\beta^0\right)\\
& +\frac{1}{n}\sum_{i=1}^{n}\left[\dot{h}_{u}(q_{0},X^{\top}_{i} \tilde{\beta})-\dot{h}_{u}(\tilde{q},X^{\top}_{i}\hat{\beta})\right]X^{\top}_i\left(\hat{\beta}-\beta^0\right)\\
&-\sum_{\{i:\delta_{i}=1\}} \hat{w}_{i}\left[\dot{h}_{u}(q_{0},X^{\top}_{i} \tilde{\beta})-\dot{h}_{u}(\tilde{q},X^{\top}_{i}\hat{\beta})\right]X^{\top}_i\left(\hat{\beta}-\beta^0\right)\\
&+ \sum_{\{i:\delta_{i}=1\}}\left(\hat{w}_{i}-w^*_i\right)\xi_{i}\left(q_{0}\right)\\
&+ \sum_{\{i:\delta_{i}=1\}}w^*_i\xi_{i}\left(q_{0}\right)\triangleq \sum^{5}_{i=1}L_{ni},
\end{split}
\end{equation}
where $\tilde{\beta}$ is between $\beta^0$ and $\hat{\beta}$. 
Note that
\begin{equation*}
\begin{split}
|L_{n1}|
\leq & \left\|\frac{1}{n}\sum_{i=1}^{n}\dot{h}_{u}(\tilde{q},X^{\top}_{i}\hat{\beta})X^{\top}_i-\sum_{\{i:\delta_{i}=1\}} \hat{w}_{i}\dot{h}_{u}(\tilde{q},X^{\top}_{i}\hat{\beta})X^{\top}_i\right\|_{\infty}\left\|\hat{\beta}-\beta^0\right\|_1.
\end{split}
\end{equation*}
By the definition of $\hat{w}$, we have
\[\left\|\frac{1}{n}\sum_{i=1}^{n}\dot{h}_{u}(\tilde{q},X^{\top}_{i}\hat{\beta})X^{\top}_i-\sum_{\{i:\delta_{i}=1\}} \hat{w}_{i}\dot{h}_{u}(\tilde{q},X^{\top}_{i}\hat{\beta})X^{\top}_i\right\|_{\infty}\leq \Delta.\]
This together with \eqref{eq: beta 1} proves $L_{n1}= O_p\left(\Delta \|\beta^0\|_0\sqrt{\log(p)/n}\right)$. By the requirements on $\Delta$ and assumptions on $\|\beta^0\|_0$, we have
\begin{equation}\label{Ln1}
L_{n1}= o_p(n^{-1/2}).
\end{equation} 
By Condition (A.3)(ii), it follows that
\begin{equation*}
\begin{split}
|L_{n2}|
\leq & c\frac{1}{n}\sum_{i=1}^{n}\left\{|\tilde{q}-q_0|+\big|X^{\top}_i\left(\hat{\beta}-\beta^0\right)\big|\right\}\big|X^{\top}_i\left(\hat{\beta}-\beta^0\right)\big| \\
\leq &c|\tilde{q}-q_0|\left\|\hat{\beta}-\beta^0\right\|_1+ c\left\|\hat{\beta}-\beta^0\right\|^2_1
\end{split}
\end{equation*}
and
\begin{equation*}
\begin{split}
|L_{n3}|
\leq & c\sum_{\{i:\delta_i=1\}}|\hat{w}_i|\left\{|\tilde{q}-q_0|+\big|X^{\top}_i\left(\hat{\beta}-\beta^0\right)\big|\right\}\big|X^{\top}_i\left(\hat{\beta}-\beta^0\right)\big|\\
\leq & c\sum_{\{i:\delta_i=1\}}|\hat{w}_i|\left(|\tilde{q}-q_0|\left\|\hat{\beta}-\beta^0\right\|_1+ \left\|\hat{\beta}-\beta^0\right\|^2_1\right).
\end{split}
\end{equation*}
 Lemma 2 in the supplementary material proves that 
 \[\tilde{q}-q_0=O_p\left(\sqrt{\left\|\beta^0\right\|_0\frac{\log(p)}{n}}\right).\] 
 These together with \eqref{eq: beta 1} and $\sum_{\{i:\delta_i=1\}}|\hat{w}_i| = O_p(1)$ prove 
 \begin{equation}\label{Ln23}
|L_{ni}|=O_p\left(\|\beta^0\|^2_0\frac{\log(p)}{n}\right)=o_p\left(n^{-1/2}\right),\quad i=2, 3.
\end{equation}
by the assumption on $\|\beta^0\|_0$. Lemma 7 in the supplementary material proves that $L_{n4} =o_p(n^{-1/2})$.
 This together with  \eqref{Ln decomp}, \eqref{Ln1} and \eqref{Ln23} proves
 \begin{equation}\label{Ln: res}
\hat{F}_n\left(q_{0}\right)-F_{n}\left(q_{0}\right)
 =\sum_{\{i:\delta_{i}=1\}}w^*_i\xi_{i}\left(q_{0}\right).
\end{equation}
Recalling the definition of $w^*$ in \eqref{eq: w star}, relations \eqref{eq: ano repre} and \eqref{Ln: res} together prove \eqref{taylor}. 

The proof of \Cref{theo: main} is then completed.
\end{proof}

\begin{proof}[Proof of \Cref{theo: main var}]
We first prove $\hat{\sigma}^2\stackrel{p}{\rightarrow}\sigma^2$. Denote 
\[V_1=\frac{1}{4}E\left[\frac{\delta\eta^{*\top}AA^{\top}\eta^*}{h(q_{0},X^{\top}\beta^0)(1-h(q_{0},X^{\top}\beta^0))}\right]\]
 and $V_2 = \Var(h(q_0,X^{\top}\beta^0))$. On the basis of Slutsky's theorem, it suffices to show that $\hat{V}_1\stackrel{p}{\rightarrow}V_1$, $\hat{V}_2\stackrel{p}{\rightarrow}V_2$ and $\hat{T}\stackrel{p}{\rightarrow}T_0$. 
 Since $\tilde{q}-q_0 = o_p(1)$ and $\|\hat{\beta}-\beta^0\|_1 = o_p(1)$, then $|\hat{T}-T_0|=o_p(1)$ and $|\hat{V}_{2}-V_2|=o_p(1)$ are proved by the law of large numbers.
$|\hat{V}_{1}-V_1|=o_p(1)$ is proved by Lemma 8 in the supplementary material. 
\end{proof} 

%
%


\bibhang=1.7pc
\bibsep=2pt
\fontsize{9}{14pt plus.8pt minus .6pt}\selectfont
\renewcommand\bibname{\large \bf References}
\expandafter\ifx\csname
natexlab\endcsname\relax\def\natexlab#1{#1}\fi
\expandafter\ifx\csname url\endcsname\relax
  \def\url#1{\texttt{#1}}\fi
\expandafter\ifx\csname urlprefix\endcsname\relax\def\urlprefix{URL}\fi

\bibliographystyle{model2-names}      
\bibliography{ref}   

 \begin{table}[htbp]
  \centering
  \caption{Bias, SD, RMSE and CP of relevant estimators based on $1000$ repetitions.}
   \setlength{\tabcolsep}{1.4mm}{
    \begin{tabular}{ccccccccccccccccccc}
    \toprule
    &	& \multicolumn{2}{c}{Bias}& \multicolumn{2}{c}{SD}& \multicolumn{2}{c}{RMSE}&\multicolumn{2}{c}{CP}\\
    	\cmidrule(lr){3-4}  \cmidrule(lr){5-6} \cmidrule(lr){7-8}\cmidrule(lr){9-10}
     &  & $\hat{q}_{aipw}$ & $\hat{q}$& $\hat{q}_{aipw}$ & $\hat{q}$& $\hat{q}_{aipw}$ & $\hat{q}$& $\hat{q}_{aipw}$ & $\hat{q}$ \\
	   \midrule
 &	& \multicolumn{7}{c}{DGP1: $\pi(X)$ is misspecified}\\
 	   \midrule	
    $n=200$ &$p=\frac{n}{4}$ &-0.241 &-0.042 &0.177 &0.196 &0.299 &0.201 &0.383 &0.952\\ 
    		   &$p=\frac{n}{2}$ &-0.300 &-0.065 &0.170 &0.206 &0.345 &0.216 &0.238 &0.941\\ 
    		   &$p=n$   &-0.338 &-0.070 &0.167 &0.212 &0.377 &0.223 &0.165 &0.933\\
		   &$p=2n$  &-0.374 &-0.104 &0.161 &0.197 &0.407 &0.222 &0.119 &0.906\\	   
    $n=400$ &$p=\frac{n}{4}$  &-0.177 &-0.027 &0.130 &0.155 &0.219 &0.158 &0.401 &0.937\\ 
    		   &$p=\frac{n}{2}$  &-0.218 &-0.039 &0.124 &0.157 &0.251 &0.162 &0.250 &0.941\\ 
    		   &$p=n$  & -0.261 &-0.048 &0.128 &0.160 &0.290 &0.167 &0.170 &0.941\\
		   &$p=2n$  &-0.133 &-0.023 &0.091 &0.114 &0.161  &0.116 &0.417 &0.941\\  
    $n=800$ &$p=\frac{n}{4}$  & -0.299 &-0.075 &0.122 &0.145 &0.323 &0.163 &0.086 &0.924\\ 
    		   &$p=\frac{n}{2}$  &-0.161 &-0.034 &0.090 &0.119 &0.185 &0.123 &0.253 &0.923\\ 
    		   &$p=n$   &-0.192 &-0.048 &0.089 &0.122 &0.211  &0.131 &0.161 &0.927\\
		   &$p=2n$  &-0.215 &-0.059 &0.084 &0.110 &0.231 &0.124 &0.088 &0.908\\	      	   
 \bottomrule
    \end{tabular}}
     \label{tab 1-1}
\end{table}%

 \begin{table}[htbp]
  \centering
  \caption{Bias, SD, RMSE and CP of relevant estimators based on $1000$ repetitions.}
  \setlength{\tabcolsep}{1.4mm}{
    \begin{tabular}{ccccccccccccccccccc}
    \toprule
    &	& \multicolumn{2}{c}{Bias}& \multicolumn{2}{c}{SD}& \multicolumn{2}{c}{RMSE}&\multicolumn{2}{c}{CP}\\
    	\cmidrule(lr){3-4}  \cmidrule(lr){5-6} \cmidrule(lr){7-8}\cmidrule(lr){9-10}
     &  & $\hat{q}_{aipw}$ & $\hat{q}$& $\hat{q}_{aipw}$ & $\hat{q}$& $\hat{q}_{aipw}$ & $\hat{q}$& $\hat{q}_{aipw}$ & $\hat{q}$ \\
	   \midrule
    &	& \multicolumn{7}{c}{DGP2: $\pi(X)$ is correctly specified}\\	
    	   \midrule	  
	$n=200$ &$p=\frac{n}{4}$  &-0.033& -0.027 &0.132 &0.130 &0.136 &0.133 &0.927 &0.936\\
    		   &$p=\frac{n}{2}$   &-0.028 &-0.017 &0.139 &0.138 &0.142 &0.139 &0.898 &0.909\\ 
    		   &$p=n$   &-0.042 &-0.025 &0.134 &0.134 &0.141 &0.137 &0.895 &0.920\\
		   &$p=2n$   &-0.043 &-0.023 &0.131 &0.130 &0.138 &0.132 &0.914 &0.931\\ 
    $n=400$ &$p=\frac{n}{4}$ &-0.014 &-0.012 &0.097 &0.097 &0.098 &0.098 &0.921 &0.911\\
    		  &$p=\frac{n}{2}$  &-0.021 &-0.017 &0.093 &0.092 &0.095 &0.093 &0.920 &0.932\\ 
    		   &$p=n$   &-0.022 &-0.015 &0.098 &0.096 &0.100 &0.098 &0.905 &0.921\\
		   &$p=2n$  &-0.031 &-0.021 &0.094 &0.095 &0.099 &0.097 &0.912 &0.915\\
    $n=800$ &$p=\frac{n}{4}$  &-0.012 &-0.012 &0.069 &0.069 &0.070 &0.070 &0.926 &0.926\\
    		  &$p=\frac{n}{2}$ &-0.013 &-0.013 &0.068 &0.067 &0.069 &0.069 &0.922 &0.921\\  
		   &$p=n$  &-0.016 &-0.015 &0.067 &0.067 &0.069 &0.068 &0.930 &0.932\\
		   &$p=2n$  &-0.023 &-0.019 &0.070 &0.069 &0.073 &0.072 &0.894 &0.897\\	      	   
 \bottomrule
    \end{tabular}}
     \label{tab 1-2}
\end{table}%

 \begin{table}[htbp]
  \centering
  \caption{Comparison of SD and ESD of the proposed estimator $\hat{q}$}
    \begin{tabular}{ccccccccccccccccc}
    \toprule
    &	&& \multicolumn{2}{c}{DGP1}&& \multicolumn{2}{c}{DGP2}\\
    	\cmidrule(lr){4-5}  \cmidrule(lr){7-8}
     &  && SD & ESD && SD & ESD \\
	   \midrule
    $n=200$ &$p=\frac{n}{4}$     && 0.196&0.204&&0.130& 0.123\\ 
    		   &$p=\frac{n}{2}$     && 0.206&0.214&&0.138& 0.122\\ 
    		   &$p=n$     &&0.212&0.220&&0.134&0.121\\
		   &$p=2n$   &&0.197&0.196&&0.130&0.121\\	   
    $n=400$ &$p=\frac{n}{2}$     && 0.155&0.152&&0.097& 0.087\\ 
    		   &$p=\frac{n}{2}$     && 0.157&0.155&&0.092& 0.087\\ 
    		   &$p=n$     &&0.160&0.162&&0.096&0.087\\
		   &$p=2n$   &&0.145 &0.145&&0.095&0.086 \\    
    $n=800$ &$p=\frac{n}{2}$     && 0.114&0.115&&0.069& 0.062\\ 
    		   &$p=\frac{n}{2}$     && 0.119&0.117&&0.067& 0.062\\ 
    		   &$p=n$     &&0.122 &0.123 &&0.067&0.062\\
		   &$p=2n$   &&0.110 &0.109 &&0.069&0.062\\	   	   
 \bottomrule
    \end{tabular}
     \label{tab 2}
\end{table}

 \begin{table}[htbp]
  \centering
  \caption{Estimates and confidence intervals (CI) of $m_1$, $m_0$ and $m$}
    \begin{tabular}{cccccccccccc}
    \toprule
     &\multicolumn{1}{c}{$m_1$} && \multicolumn{1}{c}{$m_0$} && \multicolumn{1}{c}{$m$} \\
\midrule
Estimate &308  && 260  &&48\\
CI & $(292.1,$ $323.9)$  && $(241.7,$ $278.3)$  &&$(21.6,$$74.4)$\\	   
 \bottomrule
    \end{tabular}
     \label{tab 3}
\end{table}%


\end{document}